\documentclass[twocolumn,preprintnumbers,amsmath,amssymb]{revtex4}

\def\be{\begin{equation}}
\def\ee{\end{equation}}

\def\be{\begin{equation}}
\def\ee{\end{equation}}

\def\be{\begin{equation}}
\def\ee{\end{equation}}

\usepackage[dvipdfmx]{graphicx}
\usepackage{here,braket,booktabs}
\usepackage{comment}
\usepackage{epsf}
\usepackage{amsmath}
\usepackage{graphics}
\usepackage{amsfonts}
\usepackage{amssymb}
\usepackage{latexsym}
\usepackage{color}
\input{colordvi.tex}
\usepackage{feynmp}

\begin{document}
\preprint{RESCEU-10/17}

\title{The generalized second law of thermodynamics and cosmological decoherence}

\author{Naritaka Oshita}
\affiliation{
  Research Center for the Early Universe (RESCEU), Graduate School
  of Science,\\ The University of Tokyo, Tokyo 113-0033, Japan
}
\affiliation{
  Department of Physics, Graduate School of Science,\\ The University of Tokyo, Tokyo 113-0033, Japan
}

\begin{abstract}
We pointed out that the generalized second law of thermodynamics on a de Sitter universe whose energy density
stochastically fluctuates due to quantum fluctuations is seemingly violated. We have shown that even in such a case,
the generalized second law is unviolated by taking cosmological decoherence into account.
It has been well known that the decoherence is necessary to give a reasonable reason why our universe
looks classical. Our proposal can support the importance of decoherence from another aspect, i.e. the generalized second law of thermodynamics.
\end{abstract}


\maketitle

\section{Introduction}
\begin{figure}[t]
\begin{center}
\includegraphics[keepaspectratio=true,height=40mm]{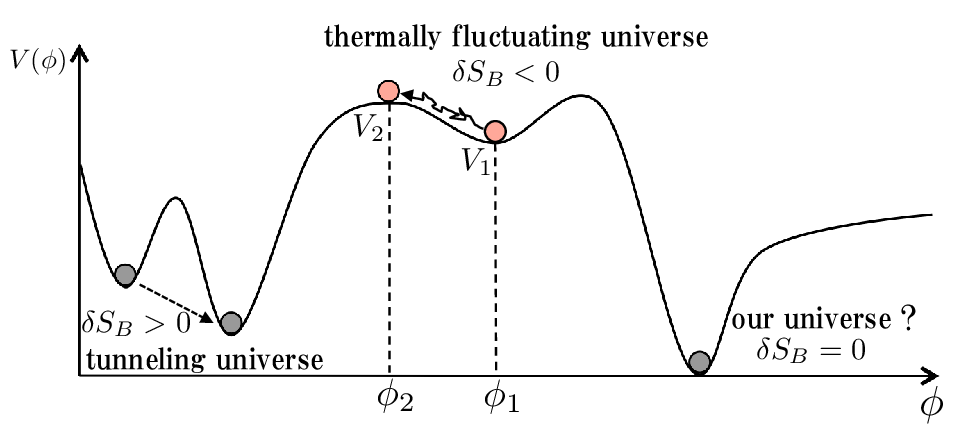}
\end{center}
\caption{The schematic picture of the landscape in which some universes
may be thermally excited ($\delta H > 0$) (red), tunnel to a more stable state ($\delta H < 0$)
or stay in a lower energy vacuum ($\delta H = 0$) (gray).
Our universe might be the latest case. 
  }%
  \label{fig1}
\end{figure}

The second law of thermodynamics states that ``\textit{the entropy of an isolated system does not decrease}" and
the Universe is an isolated system including all of the entropy in it.
The picture of the Universe has been drastically changed and the landscape \cite{Susskind:2003kw} is one of the most innovative pictures where
the spacetime accommodates a great number of universes with the various vacuum energies and our Universe is just one of them.
Although our Universe has already experienced inflation, other universes still inflate
and some of them may change their vacuum energies by thermally fluctuating on a gently curved effective potential $V(\phi)$
\cite{Linde:1986fd,Linde:1982uu} (or
by quantum tunneling a potential barrier
\cite{Coleman:1977py,Callan:1977pt,Coleman:1980aw}),
where $\phi$ is the inflaton field (Fig. \ref{fig1}).
Such a thermally fluctuating universe can be described by the stochastic inflation scheme \cite{Starobinsky:1982ee,Starobinsky:1994bd}
and this universe seemingly violates the generalized second law (GSL) of thermodynamics
\cite{Bekenstein:1972tm,Bekenstein:1973ur,Bekenstein:1974ax} as is explained below.
(The GSL is a conjecture which states that the sum of the ordinary entropy plus the Bekenstein entropy \cite{Bekenstein:1972tm,Bekenstein:1973ur}
of gravitational horizons cannot decrease in time.
It was initially formulated for black holes by Bekenstein \cite{Bekenstein:1974ax} and was extent to de Sitter universes by Davies
\cite{Davies:1987ti}.).
Let us consider a universe governed by the inflaton field $\phi$ which thermally (stochastically)
fluctuates on a gently curved region of effective potential in which
$|V''| \ll H^2 \equiv (8 \pi G /3) V$ is satisfied.
Its total entropy, $S$, is given by the sum of the Bekenstein entropy, $S_{\text{B}} \equiv A/4G$,
and that of the inflaton field, $S_{\text{M}}$, where $A \equiv 4 \pi / H^2$ is an area of the cosmological horizon \cite{Gibbons:1977mu}.
The former and latter originate from the gravitational and matter sectors of the total system respectively.
The entropy production of the inflaton should be zero $\delta S_M = 0$ as long as it evolves in a unitary fashion,
which means that the total entropy production $\delta S$ is equivalent to the difference in the Bekenstein
entropy
\begin{equation}
\delta S = \delta S_B = \delta \left( \frac{\pi}{G H^2} \right) = -\frac{2 \pi \delta H}{G H^3}.
\label{082101}
\end{equation}
According to the stochastic inflation, the universe could be thermally
excited from a lower energy density,
$V_1 \equiv V(\phi_1) \equiv 3 H_1^2/(8 \pi G)$,
to a higher energy density, $V_2 \equiv V(\phi_2) \equiv 3 H_2^2/(8 \pi G)$ (Fig. \ref{fig1}).
In the case of $\delta H \equiv H_2-H_1> 0$, using (\ref{082101}), it is found that the total entropy production
decreases and the decrement can be much larger than the unity.
Taking the parameters as, for example, $H_1 \simeq H_2 = 10^{13} \text{GeV}$
and $\delta H/ H_1 = 10^{-3}$, the entropy production is $\delta S \sim -10^{9}$.
This very large decrement of entropy implies the (seeming) violation of the GSL on a thermal universe.
Although Davies has proved that in the ``classical" level, a cosmological horizon area never decreases
($\delta S_B \geq 0$) if a cosmological fluid is subject to condition $\rho + p \geq 0$ and the scale factor $a (t)$ diverges
for $t \to \infty$ \cite{Davies:1988dk},
we here take into account the stochastic fluctuation of inflaton that is
``quantum-mechanically" driven \cite{Starobinsky:1982ee,Starobinsky:1994bd}.
Actually, Oshita and Yokoyama have shown that
\cite{Oshita:2016oqn} the cosmological horizon area may decrease due to the
Hawking-Moss transition \cite{Hawking:1981fz},
which is the quantum jump of a spatially homogeneous vacuum energy.
Here the question is whether there exist inflationary universes in which the GSL
is violated by their quantum fluctuations. If there is such a universe, it would be a
counterexample for the GSL and attributing an entropy to cosmological horizons
would be unreasonable.

In this paper, however, we will show that the GSL may be unviolated
by taking into account the decoherence of the inflaton field \cite{Polarski:1995jg,Kiefer:2008ku,Kiefer:1998qe,Campo:2005sy,Lyth:2006qz,Kiefer:1999sj,Burgess:2006jn,Kieferdeco1,Hollowood:2017bil,Kieferdeco2}
by which the entropy production takes place ($\delta S_M > 0$).
Decoherence is necessary to explain why our universe looks classical although the origin of
the structure of the Universe is considered to be quantum fluctuations, see e.g.,
\cite{Kieferdeco2}.
Our argument can also support the importance of decoherence from another aspect, i.e. the GSL.

\section{Formalism}
The stochastic inflation scheme starts with splitting the quantum fluctuations into two components,
the short-wavelength modes $\phi_>$ (sub-horizon modes) and long-wavelength modes $\phi_<$
(super-horizon modes). The former is regarded as mere white noises interacting with the latter.
Here the inflaton field $\phi$ can be described as
\begin{equation}
\phi (x) = \phi_> (x) + \phi_< (x)
\end{equation}
with
\begin{align}
\phi_> (x) &= \int_{k > \Lambda} \frac{d^3 k}{(2\pi)^3} \phi ({\bf k}, \eta) e^{i {\bf k} \cdot {\bf x}},\\
\phi_< (x) &= \int_{k < \Lambda} \frac{d^3 k}{(2\pi)^3} \phi ({\bf k}, \eta) e^{i {\bf k} \cdot {\bf x}},
\end{align}
where $\Lambda$ is a certain cutoff and $\eta$ is the conformal time.
In the Friedmann-Lema{\^i}tre-Robertson-Walker (FLRW) metric
\begin{equation}
ds^2 = a^2(\eta) \left[ d \eta^2 - d{\bf x}^2 \right],
\end{equation}
we can set the cutoff $\Lambda = \epsilon H a (\eta)$, where $a(\eta)$ is a scale factor and
$\epsilon$ is a small constant parameter \cite{Starobinsky:1994bd}.
Coarse-graining the sub-horizon modes which are inaccessible environment interacting with the
super-horizon modes leads to the decoherence and entropy production
\cite{Kiefer:1999sj,Campo:2005sy,Kiefer:2008ku,Hollowood:2017bil,Brandenberger:1992sr,Brandenberger:1992jh}, $S_M > 0$, which could offset the
decrement of the Bekenstein entropy, $S_B + S_M > 0$, as is shown in the latter part of this paper.

Let us consider a scalar field (inflaton) $\phi$ and an external scalar field (environment) $\varphi$
which interacts with $\phi$ in a curved spacetime.
The action, $S = S_{\phi} + S_{\varphi} + S_{\text{int}}$, is modeled as
\cite{Hollowood:2017bil,Boyanovsky:2015jen}
\begin{align}
\begin{split}
S_{\phi} &\equiv \int d^4x \sqrt{-g} \left[ \frac{1}{2} g^{\mu \nu} \partial_{\mu} \phi \partial_{\nu} \phi
-\frac{1}{2} m^2 \phi^2 \right],\\
S_{\varphi} &\equiv \int d^4x \sqrt{-g} \left[ \frac{1}{2} g^{\mu \nu} \partial_{\mu}
\varphi \partial_{\nu} \varphi - \frac{1}{2} \xi {\mathcal R} \varphi^2 \right],\\
S_{\text{int}} &\equiv
- \lambda \int d^4x \sqrt{-g} \phi \varphi^2,
\end{split}
\end{align}
where $m$ is the mass of $\phi$,
$\lambda$ is a coupling constant, $\xi$ is the non-minimal coupling
constant, and ${\mathcal R}$ is the Ricci scalar.
Redefining the fields as $\chi \equiv a \phi$ and $\psi \equiv
a \varphi$,
$S_{\phi}$, $S_{\varphi}$ and $S_{\text{int}}$ reduce to
\begin{align}
\begin{split}
S_{\phi} \equiv S_{\chi} &= \int d^4 x \left[ \frac{1}{2} \eta^{\mu \nu} \partial_{\mu} \chi \partial_{\nu} \chi -\frac{1}{2} M_{\chi}^2 a^2 \chi^2 \right],\\
S_{\varphi} \equiv S_{\psi} &= \int d^4 x \left[ \frac{1}{2} \eta^{\mu \nu} \partial_{\mu} \psi \partial_{\nu} \psi -\frac{1}{2} M_{\psi}^2 a^2 \psi^2 \right],\\
S_{\text{int}} &=\int d^4 x \frac{\lambda}{H \eta} \chi \psi^2,
\label{071901}
\end{split}
\end{align}
where $M_{\chi}^2 \equiv m^2 - a''/a^3$ and $M_{\psi}^2 \equiv \xi {\mathcal R}
- a''/a^3$. Here we are interested in the case of de Sitter spacetime,
and therefore the mass terms $M_{\chi}^2$ and $M_{\psi}^2$
in (\ref{071901}) reduce to $M_{\chi}^2 = m^2 -2 H^2$ and $M_{\psi}^2 = (12 \xi-2) H^2$ respectively.
We take $\psi$ to be a conformally coupled field by taking $\xi = 1/6$
($M_{\psi} = 0$).
The conformally coupled field does not feel the cosmic expansion, and therefore the field $\psi$ is not squeezed while the field $\chi$ is getting squeezed after horizon exit. Therefore, we can regard the fields $\chi$ and $\psi$ as a super-horizon and sub-horizon modes respectively.
In this sense, this model attempts to model an IR-UV split of a self-interacting single field.

Coarse-graining the environment field $\psi$, which leads to decoherence,
corresponds to taking trace over $\psi$ as
\begin{align}
\begin{split}
&\int {\mathcal D} \psi^+ {\mathcal D}\psi^- \rho(\chi^+, \chi^-, \psi^+, \psi^-; \eta) \delta (\psi^+ - \psi^-)\\
&\equiv \rho_R (\chi^+, \chi^-; \eta),
\end{split}
\end{align}
where $\rho (\chi^+, \chi^-, \psi^+, \psi^-; \eta)$ is the total density matrix and
$\rho_R (\chi^+, \chi^-; \eta)$ is the reduced density matrix.
Assuming the weak interaction between $\phi$ and $\psi$, the reduced density matrix can be factorized as
\begin{equation}
\rho_R (\chi^+, \chi^-; \eta) = \prod_{{\bf k}} \otimes \rho_R (\chi^+_{\bf k}, \chi^-_{\bf k}; \eta) + {\mathcal O} (\lambda^3),
\end{equation}
and in the limit of $|k \eta| \ll 1$, its master equation is given by \cite{Hollowood:2017bil}
\begin{align}
\begin{split}
\frac{d}{d \eta} \rho_R (\chi^+_{\bf k}, \chi^-_{\bf k}; \eta) &\simeq 
-i {\mathcal L}^{(u)}_{\bf k} [\chi_{\bf k}, \partial_{\chi_{\bf k}}] \ \rho_R (\chi^+_{\bf k}, \chi^-_{\bf k}; \eta)\\
&- \frac{\lambda^2}{8 \pi H^2 \eta^2}
|\chi^+_{\bf k} - \chi^-_{\bf k}|^2 \rho_R (\chi^+_{\bf k}, \chi^-_{\bf k}; \eta),
\label{072701}
\end{split}
\end{align}
where ${\mathcal L}^{(u)}_{\bf k}$ is the unitary time-evolution operator for the field $\chi$.
The second term in (\ref{072701}) suppresses the non-diagonal terms and leads to
decoherence. In the following, we will omit the suffix ${\bf k}$.
Solving (\ref{072701}), one obtains
\begin{align}
\begin{split}
\rho_R (\chi^+, \chi^-; \eta) &\simeq \rho_0(\chi^+, \chi^-; \eta) e^{- \frac{D (\eta)}{2} |\chi^+ - \chi^-|^2},\\
D (\eta) &\equiv - \frac{\lambda^2}{12 \pi H^2 \eta}.
\end{split}
\label{072001}
\end{align}
where $\rho_0$ is the unitary density matrix and the effect of interaction is encoded
in the exponential factor in (\ref{072001}), which suppresses the non-diagonal components of the reduced
density matrix and leads to the decoherence.

\begin{figure*}[t]
\begin{center}
\includegraphics[keepaspectratio=true,height=85mm]{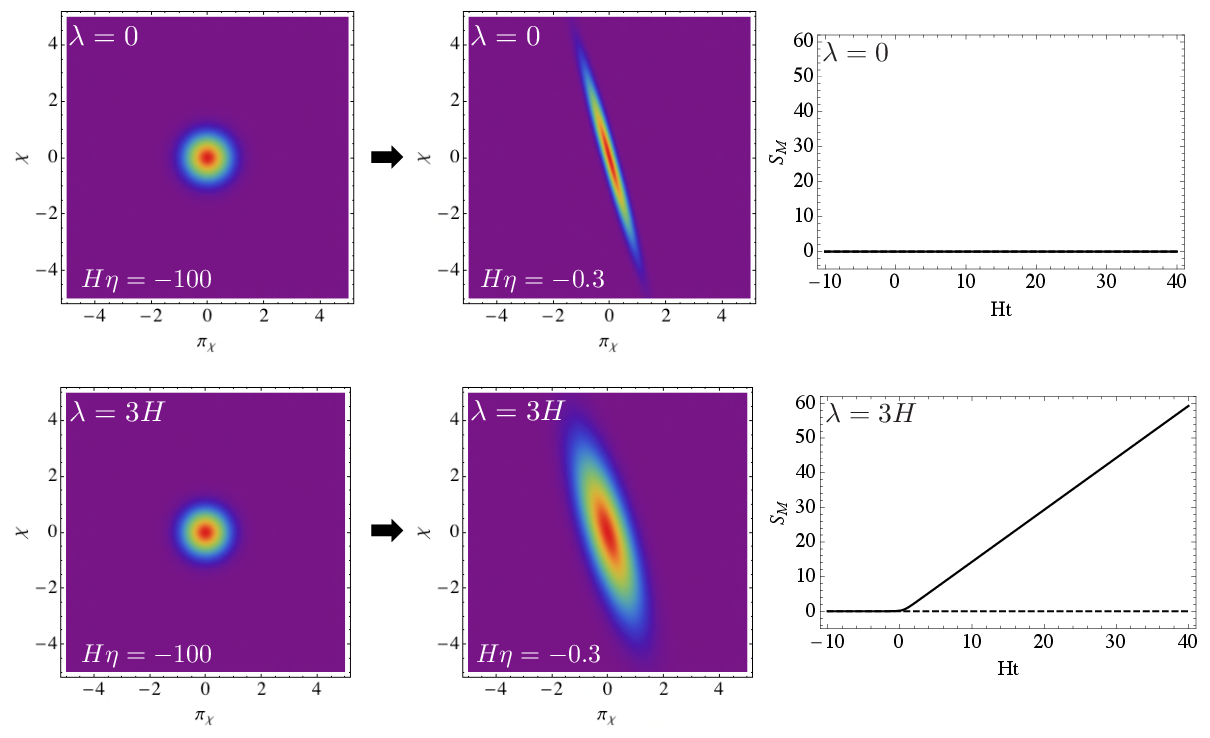}
\end{center}
\caption{
Plots of Wigner functions (left and middle) and the time evolutions of entropy (right) with $\lambda = 0$, $m=0$, $k = H$ (top)
and $\lambda = 3H$, $m=0$, $k = H$ (bottom).
}
  \label{fig2}
\end{figure*}

\section{Entropy production}
For the calculation of an entropy production $\delta S_M$, a Wigner function that is a probability distribution function on phase space is useful.
The definition of Wigner function is
\begin{align}
\begin{split}
w(\chi, \pi_{\chi}; \eta) \equiv & \frac{1}{\pi^2} \int dx_R dx_I\\
&e^{2 i \pi_{\chi} {}_R x_R+2 i \pi_{\chi} {}_I x_I} \rho (\chi+ x/2, \chi-x/2),
\end{split}
\label{072003}
\end{align}
where $\pi_{\chi}$ is the conjugate momentum of $\chi$ and the suffixes $R$ and $I$ indicate a real and imaginary part
respectively.
Taking the Bunch-Davies vacuum state, in which the mode function, $f_{\chi} (k, \eta)$, is given by
\begin{align}
\begin{split}
f_{\chi} (k, \eta) &= e^{-i3 \pi/4 +i \nu \pi / 2} \frac{\sqrt{- \pi \eta}}{2} H^{(1)} (\nu; -k \eta)\\
\text{with} \ \nu &\equiv \sqrt{\frac{9}{4} - \frac{m^2}{H^2}},
\end{split}
\end{align}
the unitary density matrix $\rho_0$ has the form \cite{Kiefer:1999sj,Hollowood:2017bil}
\begin{align}
\begin{split}
&\rho_0 (\chi^+, \chi^-; \eta) = \sqrt{ \frac{2 \Omega_R}{\pi}}\\
&\times e^{\left( -\frac{\Omega_R}{2} (\chi^+-\chi^-)^2 -i \Omega_I (\chi^+-\chi^-) (\chi^+ + \chi^-)
-\frac{\Omega_R}{2} (\chi^+ + \chi^-)^2 \right)},
\label{170418}
\end{split}
\end{align}
where $\Omega_R$ and $\Omega_I$ are the real and imaginary part of the function $\Omega (k,\eta)
\equiv -i \left( f_{\chi}^{\ast} {}'/f_{\chi}^{\ast} - a H \right)$.
From (\ref{072001}), (\ref{072003}) and (\ref{170418}), the Wigner function for the reduced density matrix
$\rho_R$ is obtained as
\begin{align}
\begin{split}
w_R (\chi,\pi_{\chi}; \eta) &= \frac{4}{\pi^2} \frac{\Omega_R}{\Omega_R + D}\\
&\times \exp{\left( -2 \frac{|\pi_{\chi} -2 \Omega_I \chi|^2}{\Omega_R + D} -2 \Omega_R |\chi|^2 \right)}.
\label{072102}
\end{split}
\end{align}

As is shown in Fig.2, one finds that the Wigner distribution is squeezed due to the cosmic expansion.
In the case of $\lambda = 0$, the system is obviously in a pure state
and the area of Wigner ellipse $A (\eta)$ remains constant, that is, the entropy production is zero $\delta S_M = 0$.
On the other hand, when $\lambda > 0$ and $m \ll H$,
smearing out the degrees of freedom of the environment $\psi$ may affect the state of $\phi$
so that its number of states (entropy) monotonically grows.
The number of states $W$ is proportional to the area of Wigner ellipse, $A \equiv \pi \alpha \beta$.
From (\ref{072102}), $A$ is given by \cite{Hollowood:2017bil}
\begin{equation}
A = \frac{\pi}{2} \left( 1 + \frac{D}{\Omega_R} \right)^{1/2}.
\end{equation}
Defining $W \equiv (2/\pi) A$ so that the matter entropy is zero when the interaction is turned off
($\lambda = 0$), the entropy is given by
\begin{equation}
S_M \equiv \ln{W} = \frac{1}{2} \ln{\left( 1 + \frac{D}{\Omega_R} \right)}
= \frac{1}{2} \ln{\left( 1 + \frac{\tilde{\lambda}^2}{12 \pi  (-H \eta) \omega_R} \right)},
\label{072703}
\end{equation}
where $\tilde{\lambda} \equiv \lambda/H$ and $\omega_R \equiv \Omega_R/H$.
In the limit of $|k \eta| \ll 1$, we found the asymptotic behavior of $\omega_R$ as
\begin{align}
\omega_R \propto
(-H\eta)^{-1+2 \nu} \ \text{for} \ m < \frac{3}{2} H.
\label{072704}
\end{align}
We now can estimate the entropy production rate due to decoherence,
$\dot{S}_M \equiv dS_M/ dt$, from (\ref{072703}) and (\ref{072704}).
For the case of $m \ll H$, the entropy production rate $\dot{S}_M$ is
\begin{equation}
\dot{S}_M \simeq \nu H = \frac{3}{2} H + {\mathcal O} (m^2/H^2).
\end{equation}
The previous works \cite{Kiefer:1999sj,Campo:2005sy,Kiefer:2008ku,Brandenberger:1992sr,Brandenberger:1992jh} in which other models are used to estimate the entropy production rate, $\delta S_M$,
also predict that the rate is of the order of Hubble parameter $\delta \dot{S}_M \sim H$.
This implies that the entropy is constantly produced with the cosmological timescale, $t \sim 1/H$,
which is caused by the squeezing \cite{Kiefer:1999sj,Brandenberger:1992sr,Brandenberger:1992jh}. Squeezing can equivalently be rephrased as ``particle creation" \cite{Kiefer:1999sj}
with average particle number $n \sim e^{2r} = e^{(2 \nu -1) Ht} \propto \omega_R^{-1}$
\cite{Brandenberger:1992sr,Brandenberger:1992jh}.
Although the created particles are in a pure state in the case of $\lambda = 0$,
once the interaction is turned on ($\lambda > 0$), particle correlation is leaked into the environment
degrees of freedom which are inaccessible and the particles apparently lose their correlation, see, e.g., \cite{Campo:2005sy}.
In this sense, we can say that the endless creation of less correlated thermal particles
may constantly produce its entropy with timescale $\sim 1/H$.

\section{GSL on thermally fluctuating universes}
Now our concern is if the entropy production $\delta S_M \sim H \delta t$ could recover the GSL
by offsetting the decrement of Bekenstein entropy $\delta S_B \sim
- \delta H/(G H^3)$.
In the first place, to observe a thermally (stochastically) excited universe, the condition
\begin{equation}
\frac{H}{2 \pi} \gg \frac{V' (\phi)}{3 H^2}
\label{072708}
\end{equation}
should be satisfied. This is because, in the stochastic inflation scheme, the coarse-grained field
$\phi$ follows the Langevin equation \cite{Starobinsky:1982ee,Starobinsky:1994bd,Vennin:2015hra}
\begin{equation}
\frac{d \phi}{d N} = - \frac{V' (\phi)}{3 H^2} + \frac{H}{2 \pi} \xi(N),
\label{072809}
\end{equation}
where $N \equiv Ht$ and $\xi (N)$ is a white noise whose origin is a quantum fluctuation,
from which one can read the condition (\ref{072708}) for a dominant thermal noise
(the second term in (\ref{072809})). 
Let us consider the situation where the field $\phi$ goes up a gentle slope of effective potential $V(\phi)$
by a step
$\delta \phi \sim H/2 \pi$ \cite{Linde:1982uu}
within the cosmological timescale $\delta t \sim 1/H$.
Replacing $V'(\phi)$ by $\delta V/ \delta \phi$ and using the Friedmann equation $V=3 H^2/(8\pi G)$,
the condition (\ref{072708}) reduces to $\delta \phi \gg \delta H / 2 G H^2$ and we obtain
\begin{equation}
1 \gg \frac{\delta H}{G H^3}.
\label{072811}
\end{equation}
Remembering $\delta S_M \sim H \delta t \sim 1$ and $\delta S_B \sim -\frac{\delta H}{G H^3}$
(see (\ref{082101})),
the inequality (\ref{072811}) reduces to the GSL:
\begin{equation}
\delta S_M + \delta S_B > 0.
\end{equation}
Now we confirm that the decoherence, which is responsible
for the entropy production $\dot{S}_M \sim H$, allows
de Sitter universes whose vacuum energy densities thermally fluctuate to be excited without the violation of the GSL.

\section{Conclusions}
In summary, using a system which models an IR-UV split of a self-interacting single field as in Refs. \cite{Hollowood:2017bil,Boyanovsky:2015jen},
we have shown that the entropy production due to the cosmological decoherence, $\delta S_M$,
could offset the decrease of the Bekenstein entropy, $\delta S_B$, during the thermal excitation
of universe. This means taking decoherence into account is necessary to satisfy
the GSL on thermal universes.
The constant entropy production $\dot{S}_M \sim H$ originates from the squeezing due to the
cosmic expansion, by which thermal particles are created and lose their quantum correlations
(i.e. quantum entanglement) due to decoherence. That is, uncorrelated thermal particles would be produced with the cosmic timescale $\sim 1/H$,
which is responsible for the entropy production $\dot{S}_M \sim H$.
Moreover, in the context of the warm inflation \cite{Berera:1995ie,Berera:2008ar} (see also \cite{Bartrum:2013fia,Bastero-Gil:2016qru})
in which the thermal equilibrium of external fields is maintained even during inflation,
the entropy production may be enhanced compared to that we have calculated.
In this sense, we have discussed if the GSL can be satisfied in a conservative setting.
Davies has shown that the GSL is satisfied in the classical level \cite{Davies:1987ti,Davies:1988dk}. On the other hand, we here have shown
that the GSL is satisfied even in the case where the quantum fluctuations of inflaton are taken into account.
This can make the GSL more reliable and can be a supporting evidence for the reconciliation between gravitation and thermodynamics.

\textit{Acknowledgements.}
This work was supported by Grant-in-Aid for JSPS Fellow No. 16J01780.
The author greatly appreciate the hospitality at Tufts university and would like to thank A.~Vilenkin and M.~Yamada for helpful discussions.

\end{document}